\documentclass[aps,nofootinbib,noshowpacs,showkeys,preprintnumbers,amsmath,amssymb]{revtex4}
\usepackage{graphicx,color}
\usepackage{graphics}
\usepackage{rotating}
\usepackage{amssymb}

\usepackage{amsfonts,amscd}
\usepackage{amsmath}
\usepackage{booktabs}

\usepackage{color}

\usepackage{epsfig}
\usepackage{dcolumn}
\usepackage{bm}
\usepackage{multirow}

\def\be{\begin{equation}}
\def\ee{\end{equation}}
\def\ba{\begin{eqnarray}}
\def\ea{\end{eqnarray}}
\def\bea{\begin{eqnarray}}
\def\eea{\end{eqnarray}}
\def\bes{\begin{subequations}}
\def\ees{\end{subequations}}
\def\bear{\begin{array}}
\def\eear{\end{array}}

\usepackage{epstopdf}

\begin{document}

\title{Origin of fermion generations from extended noncommutative geometry}

\author{Hefu Yu$^{1}$}
\author{Bo-Qiang Ma$^{1,2,3}$}
\email{mabq@pku.edu.cn}
\affiliation{$^1$ School of Physics and State Key Laboratory of Nuclear Physics and Technology, Peking University, Beijing 100871,China\\
$^2$ Collaborative Innovation Center of Quantum Matter, Beijing 100871, China\\
$^3$ Center for High Energy Physics, Peking University, Beijing 100871, China}

\date{\today}

\begin{abstract}
We propose a way to understand the 3 fermion generations by the algebraic structures of noncommutative geometry, which is a promising framework to unify the standard model and general relativity. We make the tensor product extension and the quaternion extension on the framework. Each of the two extensions alone keeps the action invariant, and we consider them as the almost trivial structures of the geometry. We combine the two extensions, and show the corresponding physical effects, i.e., the emergence of 3 fermion generations and the mass relationships among those generations. We define the coordinate fiber space of the bundle of the manifold as the space in which the classical noncommutative geometry is expressed, then the tensor product extension explicitly shows the contribution of structures in the non-coordinate base space of the bundle to the action. The quaternion extension plays an essential role to reveal the physical effect of the structure in the non-coordinate base space.
\end{abstract}

\maketitle

\section{Motivation}
\label{sec0}
The phenomenon that fermions fall into 3 generations, as an experimental fact and also as an important input of the standard model, remains a mysterious issue.
It is still obscure whether the origin of fermion generations comes from the internal nature of the standard model or other unknown principles.
A clear answer to this question can hint towards a fundamental understanding of nature. Our work is devoted to this question by showing that the fermion generations are internally determined by the standard model, when it is expressed in the framework of noncommutative geometry~\cite{nc.rev}.

The noncommutative geometry generalizes an essential aspect, the geometrization, of the general relativity to the theory for all elementary particles~\cite{nc.rev}. This turns the theory into a pure geometric scenario and gives a promising framework to unify all fundamental interactions including gravity by giving up the rule of commutativity of the space~\cite{Chamseddine:1991qh,Connes:1996gi,Chamseddine:1996zu,Chamseddine:2006ep,Chamseddine:2010ud,Connes:2006qv}. Concepts in the standard model are correspondingly turned into spectral analogues of the space, and principles of physical mechanics are reinterpreted by more specialized mathematical principles, {e.g.}, the local index theorem~\cite{nc.rev}. Although this framework is quite successful to give a promising unification of the standard model and the general relativity and shed light on an algebraic understanding of the physical concepts, the number of fermion generations is still an {\it ad hoc} input of the original theory but not a result from mathematical principles. The understanding of this input involves the essential nature of the geometry.

We provide a new way to understand the {\it ad hoc} inputs and exceptive features of a theory by extracting effects hidden in usual expressions of the theory. In particular, we consider global effects, which are usually trivial, of the manifold of the noncommutative geometry and extract the nontriviality which explains the fermion generations. This involves the tensor product extension on ingredients of the space and the quaternion extension on fields and the physical action. Each of the two extensions alone, naturally coming from the algebraic structures of the geometry, keeps the physical action invariant. We reveal in this paper the nontrivial effect by combining the two extensions in the geometric paradigm.

The combination of the tensor product extension and the quaternion extension is shown to be feasible in constructing the biframe theory~\cite{Yu:2017zbv,Wu:2015wwa,Wu:2015hoa} to characterize quantum gravity. In Ref.~\cite{Yu:2017zbv}, the tensor product extension is used to extract global effects from the geometry, and the quaternion extension is used to reveal that the non-coordinate globally flat frame in the biframe theory can come from the global effects of the geometry. In this work, we also make the tensor product extension to extract global effects from the geometry, and show that under the quaternion extension those global effects can reflect the emergence of three fermion generations and the corresponding mass relationships among those generations.

This work is arranged as follows. Section~\ref{sec1} introduces the tensor product extension on ingredients of the geometry and show that this extension is trivial in usual conditions. In Section~\ref{sec2} we give the quaternion extension on the action and show that the combination of quaternion and tensor product extensions can explain the origin of 3 fermion generations. Section~\ref{sec3} is devoted to a conclusion of this work.

\section{Tensor product extension}
\label{sec1}
The geometric space of noncommutative geometry is encoded by a spectral triple $(\mathcal{A}, \mathcal{H}, D)$, which is defined by the involutive algebra $\mathcal{A}$ represented in the Hilbert space $\mathcal{H}$ and the Dirac operator $D$ in $\mathcal{H}$~\cite{nc.rev}. In the original noncommutative geometry, the standard model is expressed in the product space $F$$\times$$M$ where $F$ is a finite noncommutative geometry and $M$ is the Riemannian manifold~\cite{nc.rev}. Some conditions are necessary
to implement the standard model in the geometry. The spectral triple should be endowed with a $\mathbb{Z}/2$-grading $\gamma$ ($\gamma^2=1$) commuting with $a\in\mathcal{A}$ and anticommuting with $D$, together with a real structure of antilinear isometry $J$~\cite{nc.rev,Connes:1996gi,Chamseddine:1991qh}. The order zero condition
\begin{equation} \label{order0}
[a, b^0]=0, \quad \forall a \in \mathcal{A}, \,\,\, b^0 \in \mathcal{A}^0=\{c^0|c^0:=Jc^{*}J^{-1},~ c\in \mathcal{A}\},
\end{equation}
and the order one condition
\begin{equation} \label{order1}
[[D,a], b^0]=0, \quad \forall a \in \mathcal{A}, \,\,\, b^0 \in \mathcal{A}^0,
\end{equation}
are supposed to hold in the geometry~\cite{nc.rev}. The $K$-theoretic dimension of the product space is 10 modulo 8 to overcome the fermion doubling problem~\cite{Lizzi:1996vr,Chamseddine:1996zu}.

The geometries of $F$ are classified in~\cite{concep,why.sm}. With the symplectic assumption in the minimal nontrivial case~\cite{concep,why.sm}, the $\mathbb{Z}/2$-grading $\gamma_F$ and the order zero condition of $F$ restrict the solution of the irreducible pair $(\mathcal{A}, J)_F$ to
\begin{eqnarray}
\mathcal{A}_F &=& M_2({\mathbb{H}}) \oplus  M_4(\mathbb{C}), \label{Af.0.0} \\
J_F(x,y) &=& (y^*, x^*), \quad \mathrm{with} \quad x \in M_2({\mathbb{H}}), \,\,\, y\in M_4(\mathbb{C}). \label{Jf.0.0}
\end{eqnarray}
The nontrivial grading $\gamma_F$ breaks $M_2(\mathbb{H})$ of (\ref{Af.0.0}) into $\mathbb{H}\oplus \mathbb{H}$. The order one condition (\ref{order1}) breaks $\mathbb{H}\oplus \mathbb{H}$ into $\mathbb{C} \oplus \mathbb{H}$ (while the deduced $\mathbb{C}$ and $\mathbb{H}$ are denoted by $\mathbb{H}_R$ and $\mathbb{H}_L$ in the following) and breaks $M_4(\mathbb{C})$ to $\mathbb{C} \oplus M_3(\mathbb{C})$. As in the original work~\cite{nc.rev}, the representations of $M_3(\mathbb{C})$, $ \mathbb{H}_{LR}$, and $\mathbb{C}$ are denoted by ${\bf 3}$, ${\bf 2_{LR}}$, and ${\bf 1}$. As a result of the classification of geometries~\cite{why.sm,concep}, one can write $\mathcal{A}_F$ as
\begin{equation} \label{Af.0.1}
\mathcal{A}_F=({\bf 2}_L \oplus {\bf 2}_R) \oplus ({\bf 1}\oplus {\bf 3}),
\end{equation}
and the Hilbert space $\mathcal{H}_F$ of $\mathcal{A}_F$ is~\cite{nc.rev}
\begin{equation}\label{Hf.0}
  \mathcal{H}_F = ({\bf 2}_L \oplus {\bf 2}_R)\otimes ({\bf 1}^0 \oplus {\bf 3}^0)  \oplus ({\bf 1}\oplus {\bf 3})\otimes ({\bf 2}_L^0 \oplus {\bf 2}_R^0).
\end{equation}
Elements in $\mathcal{H}_F$ can be representations of quarks $q=|\!\! \uparrow\downarrow\rangle_{LR}\otimes {\bf 3}^0 = {\bf 2}_{LR}\otimes {\bf 3}^0$ and leptons $l=|\!\! \uparrow\downarrow\rangle_{LR}\otimes {\bf 1}^0 = {\bf 2}_{LR}\otimes {\bf 1}^0$ and the corresponding anti-particles.

In this work, we consider the connection in general relativity as the concept coming from the transformation of coordinate frames, and consider the gauge fields in the standard model as the concepts coming from the covariant derivative with gauge symmetries. Then there is a space where all the variables in the classical noncommutative geometry can be defined, and we let such coordinate space be a fiber space of a fiber bundle of the manifold. Variables in such fiber space characterize the standard model and gravity in the frame of classical noncommutative geometry, and nontrivially couple with Dirac operator of the spectral triple in the action. We define the non-coordinate base space of the bundle as the space with fields which always trivially couple with the Dirac operator. Variables in the fiber bundle can be generally written as $\alpha(x)\otimes \alpha(1)\oplus \alpha(1)\otimes \alpha(x)$, where the part restricted in the coordinate fiber space are denoted as $\alpha(x)$ and those restricted in the non-coordinate base space are denoted as $\alpha(1)$. Then we let the Dirac operator coupling with $\alpha(x)$ and $\alpha(1)$ be $D_F$ (the Dirac operator of space $F$ in the original geometry~\cite{nc.rev}) and ``1'',  respectively. Furthermore, we distinguish the scalar multiplications on fields in the base space and that on fields in the fiber. Namely, we let $k_F$ and $k_B$ be the scalars in the fiber and the base space, respectively, and we have
\begin{eqnarray}
\!\!\!\!\!\! && k_F (\alpha(x)\otimes \alpha(1)\oplus \alpha(1)\otimes \alpha(x)) = (k_F \alpha(x))\otimes \alpha(1) \oplus \alpha(1)\otimes(k_F \alpha(x)), \label{kF.alpha} \\
\!\!\!\!\!\! &&  k_B (\alpha(x)\otimes \alpha(1)\oplus \alpha(1)\otimes \alpha(x)) = \alpha(x)\otimes(k_B \alpha(1)) \oplus(k_B \alpha(1))\otimes \alpha(F), \label{kB.alpha}
\end{eqnarray}
Thus the parts of fields of the bundle restricted to the base space and the fiber of the bundle are alternatively defined to be linear structures. The algebraic axioms in the linear structures of the original noncommutative geometry can be directly used in our structures.

We extend the Dirac operator of $F$ to
\begin{equation}\label{D.1}
  \triangle D_F=\frac{1}{2}(D_F\otimes 1 + 1\otimes D_F).
\end{equation}
As mentioned above, we define fields nontrivially coupled with $D_F$ of $\triangle D_F$ as those restricted to the coordinate fiber space of the bundle. The fields coupling with ``1'' of $\triangle D_F$ are those restricted to the non-coordinate base space. Other ingredients of the spectral triple of $F$ are extended by
\begin{eqnarray}
&&  \triangle a_F = a_{F} \otimes a_{F}, \forall a_F \in \mathcal{A}_F, \label{af.1} \\
&& \triangle \gamma_F=\frac{1}{2}(\gamma_F\otimes 1 + 1\otimes \gamma_F), \label{gammaf.1} \\
&& \triangle J_F=J_F\otimes J_F. \label{Jf.1}
\end{eqnarray}
Note that from (\ref{gammaf.1}) $\triangle\gamma_F$ does not give a $\mathbb{Z}/2$-grading on Hilbert space of $\mathcal{A}_F$. In this extension, we restrict $\triangle \mathcal{A}_F$ to $\triangle \mathcal{A}_F^+=\{\triangle a_F \,|\,(\gamma_F\otimes1)\triangle a_F =(1\otimes\gamma_F)\triangle a_F, \,\, a_F\in \mathcal{A}_F\}$. Then $\triangle \gamma_F$ on $\triangle\mathcal{A}_F^+$ is a $\mathbb{Z}/2$-grading and can give the right chirality for particles represented by $\triangle \mathcal{H}_F$ and inner fluctuations of $D_F$ in later discussion. We mention that elements of $\triangle A_F^+$ are those in $\triangle A_F$ having two same subscripts ``$L$'' or ``$R$'', and such elements can form an algebra. In what follows we simply denote $\mathcal{A}_F^{+}$ by $\mathcal{A}_F$. One can safely make this simplification on the notion of $\mathcal{A}_F^{+}$ because the Hilbert space $\mathcal{H}_F^{+}=\{\xi\,|\,\gamma \xi=\xi , \xi\in \mathcal{H}_F\}$, which contains the representations of fermions in the original geometry~\cite{nc.rev,Chamseddine:2006ep,Connes:2006qv,Chamseddine:2010ud,Connes:1996gi}, and other physical notions keep unchanged under this simplification of notation. In this sense, $\triangle\gamma_F$ with (\ref{gammaf.1}) is still a $\mathbb{Z}/2$-grading of the geometry with $\mathcal{A}_F$ restricted as above.

The left-right $\mathcal{A}_F$-module is extended by
\begin{equation}\label{comod}
  \rho(\xi)= a_{F1} \otimes \xi \otimes a_{F2}^0, \quad a_{F1}, a_{F2} \in \mathcal{A}_F, \,\,\, \xi \in \mathcal{M}.
\end{equation}
$\mathcal{A}_F$ acts on $\rho(\mathcal{M})$ via
\begin{equation}\label{a.Hf}
  \triangle a_F \rho(\xi) = (a_{F1}\otimes a_{F1})(a_{F2}\otimes \xi \otimes a_{F3}^0)=(a_{F1}a_{F2})\otimes(a_{F1} \xi) \otimes a_{F3}^0,
\end{equation}
and $\mathcal{A}_F^0$ acts on $\rho(\mathcal{M})$ via
\begin{equation}\label{a.Hf}
  \triangle a_F^0 \rho(\xi) = (a_{F1}^0\otimes a_{F1}^0)(a_{F2}\otimes \xi \otimes a_{F3}^0)=a_{F2}\otimes(\xi a_{F1}^0) \otimes (a_{F3}^0a_{F1}^0).
\end{equation}
Replacing $\mathcal{H}_F$ by $\rho(\mathcal{H}_F)$ means that representations of spinors are contained by $\rho(\mathcal{H}_F)$ in our extension. We mention that $\mathcal{H}_F$ is introduced to contain ``half-spin representations'' by fulfilling the condition that the adjoint action of $s=(1,$$-1,$$-1,$$1)$ on elements of $\mathcal{H}_F$ gives ``$-1$''~\cite{nc.rev}. Here $s=(\lambda, q_L, q_R, m)=(1,$$-1,$$-1,$$1)$ with $\lambda\in \mathbb{C}$, $q_L\in \mathbb{H}_L$, $q_R\in\mathbb{H}_R$, and $m\in M_3(\mathbb{C})$. We extend $s$ to $\triangle s$ and let the the adjoint action of $\triangle s$ on $\rho(\xi)$ ($\xi\in \mathcal{H}_F$) be $-\rho(\xi)$. The tensor product extension on $s$ is defined as
\begin{equation}\label{co.s}
  \triangle s=\frac{1}{2}(s\otimes 1 + 1\otimes s).
\end{equation}
Then $\rho(\mathcal{H}_F)$ in our extension can be written as
\begin{eqnarray}
\rho(\mathcal{H}_F) &=& ({\bf 2}_L\otimes{\bf 2}_L \oplus {\bf 2}_R\otimes{\bf 2}_R) \otimes ({\bf 3}^0\oplus {\bf 1}^0)\otimes({\bf 3}^0\oplus {\bf 1}^0)\oplus \mathrm{h.c.} \nonumber \\
   &=& \theta_{1L} \otimes \theta_{1L} \oplus \theta_{1R} \otimes \theta_{1R}  \oplus \theta_{1L}\otimes \theta_{2L} \oplus \theta_{1R}\otimes \theta_{2R} \nonumber \\
  & & \oplus \theta_{2L} \otimes \theta_{1L} \oplus \theta_{2R} \otimes \theta_{1R} \oplus \theta_{2L} \otimes \theta_{2L} \oplus \theta_{2R} \otimes \theta_{2R}  \oplus \mathrm{h.c.,} \label{Hf}
\end{eqnarray}
where $\theta_1 = {\bf 2}\otimes{\bf 3}^0$, $\theta_2 = {\bf 2}\otimes{\bf 1}^0$, $\bar{\theta}_1 = {\bf 3}\otimes{\bf 2}^0$, and $\bar{\theta}_2 = {\bf 1}\otimes{\bf 2}^0$.

As mentioned above, we define factors of $\triangle a_F$ and $\rho(\mathcal{H}_F)$ nontrivially coupling with $D_F$ in $\triangle D_F$ as the fields restricted to the coordinate fiber space of the bundle, and define the factors trivially coupling with the ``1'' of $\triangle D_F$ as the fields restricted to the non-coordinate base space of the bundle. Then we explicitly rewrite $\triangle a_F$ and $\rho(\mathcal{H}_F)$ as, in the mentioned form of tensor products of $\alpha(x)$ and $\alpha(1)$,
\begin{eqnarray}
  \triangle a_F &=& a_F(x)\otimes a_F(1) \oplus a_F(1)\otimes a_F(x), \label{a.x1} \\
  \rho(\mathcal{H}_F) &=& \sum_{i,j,Z} (\theta_{iZ}(x)\otimes \theta_{jZ}(1)\oplus \theta_{iZ}(1)\otimes \theta_{jZ}(x)\oplus \mathrm{h.c.}), \label{h.x1}
\end{eqnarray}
where $a_F(x)$ and $\theta(x)$ are the fields restricted in the coordinate fiber space and $a_F(1)$ and $\theta(1)$ are those restricted to the non-coordinate base space of the bundle, and we summate over $i,j \in \{1,2\}$, $Z\in\{L,R\}$ in (\ref{h.x1}).

We let the generation number of fermions represented by fields in the coordinate fiber space fulfills the minimal condition, i.e., the fermions in the coordinate fiber space have 1 generation. Then we show that in the total space there are 3 fermion generations by combining our tensor product extension and the quaternion extension in Section \ref{sec2}. The tensor product extension explicitly contributes extra factors from the non-coordinate structure. These factors do not have physical effect in usual conditions. However, as shown in Section \ref{sec2}, these extra factors combined with the quaternion extension play an essential role in the emergence of 3 fermion generations.

We mention that our extension does not map space $F$ to $F$$\times$$F$. The extension just maps ingredients of the geometry of $F$ to other expressions, keeping the $K$-theoretic dimension of $F$ as 6 modulo 8 to overcome the fermion doubling problem~\cite{Lizzi:1996vr,Chamseddine:1996zu}, and thus the conditions
\begin{equation} \label{JDgamma}
(\triangle J_F)^2=1,\,\,\, \triangle J_F \triangle D_F= \triangle D_F\triangle J_F,\,\,\, \triangle J_F \triangle\gamma_F=-\triangle\gamma_F\triangle J_F,
\end{equation}
determined by the $K$-theoretic dimension are fulfilled. One can directly check that the order zero condition
\begin{equation}\label{order0.1}
  [\triangle a_F, \triangle(b_F^0)]=0, \quad \forall a \in \mathcal{A}_F, \,\,\, b_F^0 \in \mathcal{A}_F^0,
\end{equation}
and the order one condition
\begin{equation}\label{order1.1}
  [[\triangle D_F,\triangle a_F], \triangle (b_F^0)]=0, \quad \forall a_F \in \mathcal{A}_F, \,\,\, b_F^0 \in \mathcal{A}_F^0,
\end{equation}
of extended notations of $F$ still hold.

The order one condition is not used in recent works~\cite{Chamseddine:2013kza,Chamseddine:2013rta}, in which the Pati-Salam unification~\cite{pati.salam} emerges. $\mathcal{A}_F$ selected by the classification of geometries without the order one condition is~\cite{Chamseddine:2013kza, Chamseddine:2013rta,Chamseddine:2014nxa,Chamseddine:2014uma}
\begin{equation}\label{Af.no1.0}
  M_2(\mathbb{H})\oplus M_4(\mathbb{C}).
\end{equation}
In our extension, when the order one condition is not used,
\begin{equation} \label{Af.no1.1}
\triangle \mathcal{A}_F= M_2(\mathbb{H})\otimes M_2(\mathbb{H}) \oplus M_4(\mathbb{C})\otimes M_4(\mathbb{C}).
\end{equation}
$\mathcal{A}_F$ in form (\ref{Af.0.1}) or (\ref{Af.no1.0}) gives the same result in this work. We mainly discuss the case that $\mathcal{A}_F$ is in form (\ref{Af.0.1}) under the order one condition. The discussion on the case without the order one condition can exactly follow the steps of the case under that condition.

For convenience of expressing physical notions in the extension, we rewrite the spectral triple of the product space $M\times F$ as
\begin{equation}\label{M.F}
  (\mathcal{A}, \mathcal{H}, D, J, \gamma)=(C^\infty(M), L^2(M,S), \partial\!\!\!/_M, J_M, \gamma_5)\otimes(\triangle\mathcal{A}_F, \rho(\mathcal{H}_F), \triangle D_F, \triangle J_F, \triangle \gamma_F),
\end{equation}
with
\begin{eqnarray}
  \mathcal{A} &=& C^\infty(M)\otimes \triangle\mathcal{A}_F, \label{A.1} \\
  \mathcal{H} &=& L^2(M,S)\otimes \rho(\mathcal{H}_F), \label{H.1} \\
  D &=& \partial\!\!\!/_M\otimes 1 + \gamma_5\otimes \triangle D_F, \label{D.1} \\
  \gamma &=& \gamma_5\otimes \triangle \gamma_F, \label{gamma.1} \\
  J &=& J_M \otimes \triangle J_F. \label{J.1}
\end{eqnarray}
Then we can discuss representations of particles and the physical action functional.

The representations of bosons in noncommutative geometry are derived from the inner fluctuation of the metric~\cite{Chamseddine:1991qh,Chamseddine:2010ud}. The inner fluctuation comes from the Morita equivalence~\cite{Rieffel.morita} of an algebra to itself, and replaces $D$ by
\begin{equation} \label{Da}
D_A=D+A+JAJ^{-1},
\end{equation}
where $A=\sum_i a_i [D, b_i]$ ($a_i, b_i \in \mathcal{A}$) plays a role of the gauge potential~\cite{Chamseddine:1991qh,nc.rev,Connes:2006qj}. Since $D=\partial\!\!\!/_M\otimes1+\gamma_5\otimes D_F$~\cite{Chamseddine:1991qh,Chamseddine:2010ud}, $A$ is separated to the continuous part $A^{(1,0)}$ coming from $\partial\!\!\!/_M\otimes1$ and the discrete part $A^{(0,1)}$ coming from $\gamma_5\otimes D_F$. $D_A$ is correspondingly separated into continuous part $D^{(1,0)}$ and discrete part $D^{(0,1)}$. The continuous gauge potential $A^{(1,0)}$ contains representations of gauge bosons, and the discrete gauge potential $A^{(0,1)}$ contains the representations of Higgs bosons~\cite{Chamseddine:1991qh,Connes:2006qj}.

In our tensor product extension, the continuous part of the inner fluctuation is rewritten as
\begin{equation}\label{A10.1}
  A^{(1,0)}=\sum_i a_i[\partial\!\!\!/_M\otimes 1 \otimes 1, a_i'],
\end{equation}
where $a_i=\eta\otimes \triangle a_{Fi}=\eta\otimes (\triangle \lambda_i, \triangle q_i, \triangle m_i)$ and $a_i'=\eta\otimes \triangle a_{Fi}'=\eta'\otimes(\triangle \lambda_i', \triangle q_i', \triangle m_i')$ are elements of $\mathcal{A}=C^\infty(M, \triangle \mathcal{A}_F)$, with $\eta, \eta'\in C^\infty(M)$. Since the gauge potentials and connections are restricted in the coordinate fiber space of the bundle, we can rewrite the local gauge transformations $U$ in the extended geometry as $U(x)\otimes 1\oplus 1\otimes U(x)$. $A^{(1,0)}$ then reads
\begin{eqnarray}
A^{(1,0)} &=&  \sum_{i}  \eta\otimes a_{Fi}\otimes a_{Fi}[\partial\!\!\!/_M \otimes 1 \otimes 1, \eta'\otimes a_{Fi}'\otimes a_{Fi}'] \nonumber \\
&=& \sum_{i}  ( a_{Fi}(x) [\partial\!\!\!/, a_{Fi}'(x)] \otimes a_{Fi}(1)a_{Fi}'(1) \nonumber \\
& & \oplus a_{Fi}(1)a_{Fi}'(1)\otimes a_{Fi}(x)[\partial\!\!\!/, a_{Fi}'(x)])  \nonumber \\
&=&  (A^{(1,0)}_0\otimes G) \oplus (G \otimes A^{(1,0)}_0),  \label{A10.2}
\end{eqnarray}
with
\begin{equation}\label{G}
  G=\sum_i a_{Fi}(1)a_{Fi}'(1), \quad \forall a_{Fi}, a_{Fi}' \in \mathcal{A}_F,
\end{equation}
and
\begin{equation}\label{A10.0}
  A^{(1,0)}_0=\sum_i a_{Fi}(x)[\partial\!\!\!/_M, a_{Fi}(x)'].
\end{equation}
Here $A^{(1,0)}_0$ is the continuous part of the gauge potential in the unextended geometry. As shown in~\cite{nc.rev,Chamseddine:2006ep,Connes:2006qv,Chamseddine:2010ud,Connes:1996gi}, $A^{(1,0)}_0$ corresponds to the gauge bosons of the symmetry group SU$(3)$$\times$SU$(2)$$\times$U$(1)$ with the proper symmetry charges in the standard model.

The discrete part of inner fluctuation in our tensor product extension is rewritten as
\begin{eqnarray}
  A^{(0,1)} &=& \sum_i a_{i} [\gamma_5 \otimes \triangle D_F, a_{i}'] \nonumber \\
   &=& \frac{1}{2} \sum_{i} \gamma_5 \otimes ( a_{Fi}(x)[D_F, a_{Fi}(x)']\otimes a_{Fi}(1)a_{Fi}'(1) \nonumber \\
   & & + a_{Fi}a_{Fi}'(1)\otimes a_{Fi}(x)[D_F,a_{Fi}'(x)]) \nonumber \\
   &=& \frac{1}{2} (A^{(0,1)}_0 \otimes G + G \otimes A^{(0,1)}_0), \label{A01.1}
\end{eqnarray}
where $a_i=\eta\otimes \triangle a_{Fi}$ and $a_i'=\eta\otimes \triangle a_{Fi}'$ are elements of $\mathcal{A}=C^\infty(M, \triangle \mathcal{A}_F)$, and
\begin{equation}\label{A01.0}
  A^{(0,1)}_0=\sum_i a_{Fi}(x)[D_F, a_{Fi}(x)'].
\end{equation}
$A^{(0,1)}_0$ in (\ref{A01.0}) is formally the same as that in the unextended geometry~\cite{nc.rev,Chamseddine:2006ep,Connes:2006qv,Chamseddine:2010ud,Connes:1996gi}. As shown in~\cite{nc.rev,Chamseddine:2006ep,Connes:2006qv,Chamseddine:2010ud,Connes:1996gi}, given $a_{Fi}=(\lambda_i, q_{iL}, m_i)$ and $a_{Fi}'=(\lambda_i', q_{iL}', m_i')$ with
\[m=\left(\begin{array}{cc}
\alpha & -\bar{\beta} \\
\beta & \bar{\alpha}
\end{array}
\right),\]
$A^{(0,1)}_0$ is parameterized by the vector
\begin{equation} \label{phi1}
\phi_1 = \left( \begin{array}{c} \phi_{11} \\ \phi_{12} \end{array} \right)=\left( \begin{array}{c} \sum_i {\lambda_{1i} (\alpha_{2i} - \lambda_{2i})} \\ \sum_i {\lambda_{1i} \beta_{1i}} \end{array} \right).
\end{equation}
The Higgs doublet is represented as $H=\phi_{11} + \phi_{12}j$~\cite{nc.rev}.

The contribution of $D^{(0,1)}$ to the spectral action is determined by Tr$((D^{(01)})^2)$ and Tr$((D^{(01)})^4)$~\cite{nc.rev}. For the calculation of Tr$((D^{(01)})^2)$ and Tr$((D^{(01)})^4)$, we recall that in the unextended geometry $A^{(0,1)}_0=A^{(0,1)}_q \oplus A^{(0,1)}_l$~\cite{nc.rev,Chamseddine:2006ep,Connes:2006qv,Chamseddine:2010ud,Connes:1996gi} where $A^{(0,1)}_l$ and $A^{(0,1)}_q$ respectively act on $\mathcal{H}_{fl} = {\bf 2}_R \otimes{\bf 1}^0 \oplus {\bf 2}_L \otimes{\bf 1}^0$ and $\mathcal{H}_{fq} = {\bf 2}_R \otimes{\bf 3}^0 \oplus {\bf 2}_L \otimes{\bf 3}^0$. Then we recall that in the unextended case $D_F$ acts on ${\bf 2}\otimes {\bf 3}^0$ and ${\bf 2}\otimes {\bf 1}^0$ by~\cite{nc.rev,Chamseddine:2006ep,Connes:2006qv,Chamseddine:2010ud,Connes:1996gi}
\begin{equation} \label{d0.fenkuai}
D = \left( \begin{array}{cc}
S & T*\\
T & \bar{S}
\end{array}
\right),
\end{equation}
with basis of ${\bf 2}\otimes {\bf 3}^0$ and ${\bf 2} \otimes {\bf 1}^0$ expressed as $(|\!\uparrow\rangle_{R}\otimes{\bf 3}^0, |\!\downarrow\rangle_{R}\otimes{\bf 3}^0, |\!\uparrow\rangle_{L}\otimes{\bf 3}^0, |\!\downarrow\rangle_{L}\otimes{\bf 3}^0)$ and $(|\!\uparrow\rangle_{R}\otimes{\bf 1}^0, |\!\downarrow\rangle_{R}\otimes{\bf 1}^0, |\!\uparrow\rangle_{L}\otimes{\bf 1}^0, |\!\downarrow\rangle_{L}\otimes{\bf 1}^0)$. The block $S$ is the linear map as~\cite{nc.rev}
\begin{equation} \label{s0}
S=S_l \oplus (S_q \otimes 1_3),
\end{equation}
where $S_l$ acts on ${\bf 2} \otimes {\bf 1}^0$ by
\begin{equation} \label{Sl0}
    S_l = \left( \begin{array}{cccc}
    0&0&\Upsilon_\nu^*&0 \\
    0&0&0&\Upsilon_e^* \\
    \Upsilon_\nu &0&0&0 \\
    0&\Upsilon_e&0&0
    \end{array}
    \right),
    \end{equation}
and $S_q$ acts on ${\bf 2} \otimes {\bf 3}^0$ by
    \begin{equation} \label{Sq0}
    S_q = \left( \begin{array}{cccc}
    0&0&\Upsilon_u^*&0 \\
    0&0&0&\Upsilon_d^* \\
    \Upsilon_u &0&0&0 \\
    0&\Upsilon_d&0&0
    \end{array}
    \right).
    \end{equation}
We mention that $\Upsilon_\nu$, $\Upsilon_e$, $\Upsilon_u$, and $\Upsilon_d$ in the original geometry~\cite{nc.rev} are complex 3$\times$3 matrices corresponding to the 3 generations, while in our extended geometry $\Upsilon_\nu$, $\Upsilon_e$, $\Upsilon_u$, and $\Upsilon_d$ are 1$\times$1 matrices (complex numbers). Given $a_{Fi}=(\lambda_i, q_{iL}, m_i), a_{Fi}'=(\lambda_i', q_{iL}', m_i')$ as above and the vectors $\phi_1, \phi_2$ expressed as (\ref{phi1}) and
\begin{equation} \label{phi2}
\phi_2 = \left( \begin{array}{c} \phi_{21} \\ \phi_{22} \end{array} \right)= \left( \begin{array}{c} \sum_i{\alpha_{1i} (\lambda_{2i} - \alpha_{2i})+\beta_{1i} \bar{\beta}_{2i} } \\ \sum_i \left( -\alpha _{1i} \beta_{2i} +
\beta_{1i}(\bar{\lambda}_{2i} - \bar{\alpha}_{2i}) \right) \end{array} \right),
\end{equation}
one can check that
\begin{equation} \label{a01q}
   A^{(0,1)}_q=\left( \begin{array}{cc}
   0&X_1 \\
   X_2&0
   \end{array}
   \right)\otimes 1_3,
\end{equation}
with
\begin{equation} \label{x.12}
X_1 = \left( \begin{array}{cc} \Upsilon_u^* \phi _{11} & \Upsilon_u^* \phi_{12} \\ -\Upsilon_d^* \bar{\phi}_{12} & \Upsilon_d^* \bar{\phi}_{11} \end{array} \right), \quad
X_2 = \left( \begin{array}{cc} \Upsilon_u \phi _{21} & \Upsilon_d \phi_{22} \\ -\Upsilon_u \bar{\phi}_{22} & \Upsilon_d \bar{\phi}_{21} \end{array} \right),
\end{equation}
and
\begin{equation} \label{a01l}
    A^{ (0,1)}_l=\left( \begin{array}{cc}
    0&Y_1 \\
    Y_2&0
    \end{array}
    \right),
\end{equation}
with
\begin{equation} \label{y1}
Y_1 = \left( \begin{array}{cc} \Upsilon_{\nu}^* \phi _{11} & \Upsilon_{\nu}^* \phi_{12} \\ -\Upsilon_e^* \bar{\phi}_{12} & \Upsilon_e^* \bar{\phi}_{11} \end{array} \right), \quad
Y_2 = \left( \begin{array}{cc} \Upsilon_{\nu} \phi _{21} & \Upsilon_e \phi_{22} \\ -\Upsilon_{\nu} \bar{\phi}_{22} & \Upsilon_e \bar{\phi}_{21} \end{array} \right).
\end{equation}
Then $A^{(0,1)}_q$ and $A^{(0,1)}_l$ are formally the same as those in the unextended geometry.

Now we proceed to the calculation of the contribution of $D^{(0,1)}$ to the action. We assume that $|G|^2=1$. Without this assumption, one only needs a reparameterization of the coefficients of the action to get the same result of this work. Let $X_1$ and $X_2$ in (\ref{a01q}) be equal to $X$, then Tr$((A_q^{(0,1)}\otimes G)^2)=$Tr$((A_q^{(0,1)})^2)$Tr$(|G|^2)=6$Tr$(XX^*)$Tr$(|G|^2)=6|H^2|(\Upsilon_u^*\Upsilon_u+\Upsilon_d^*\Upsilon_d)$ (with $|G|^2=1$). Similarly, one can check that Tr$((A_l^{(0,1)})^2)=2|H^2|(\Upsilon_\nu^*\Upsilon_\nu+\Upsilon_e^*\Upsilon_e)+(\Upsilon_R^*\Upsilon_R).$ By taking into account the terms $JA^{(0,1)}J^{-1}$ and $D$ in $D^{(0,1)}$, the above results of $A_q^{(0,1)}$ and $A_l^{(0,1)}$ are then multiplied by 2 and $|H|$ is replaced by $|H+1|$. Thus, we have
\begin{equation} \label{D01.2.1}
\mathrm{Tr}((D^{(01)})^2)= 4a|1+ H|^2 + 2c,
\end{equation}
where $a=\Upsilon_\nu^*\Upsilon_\nu+\Upsilon_e^*\Upsilon_e+3(\Upsilon_u^*\Upsilon_u+\Upsilon_d^*\Upsilon_d)$ and $c=\Upsilon_R^*\Upsilon_R$. Similarly, one can check that
\begin{equation} \label{D01.4.1}
\mathrm{Tr}((D^{(01)})^4) = 4b |1+ H|^4 + 2d +8e |1+H|^2,
\end{equation}
where $b=(\Upsilon_\nu^*\Upsilon_\nu)^2 +(\Upsilon_e^*\Upsilon_e)^2 +3((\Upsilon_u^*\Upsilon_u)^2 +(\Upsilon_d^*\Upsilon_d))^2$, $d=(\Upsilon_R^*\Upsilon_R)^2$, and $e=\Upsilon_R^*\Upsilon_R\Upsilon_\nu^*\Upsilon_\nu$. The coefficients $a$, $b$, $c$, $d$ and $e$ are formally the same as those in the unextended geometry, only with the notation ``Tr'' in the unextended case removed in our extension. In fact, ``Tr'' does not play a role because the variables in these coefficients are 1$\times$1 complex matrices.

From the above calculations on inner fluctuations of $D$, the spectral action~\cite{Connes:1988ym, Chamseddine:1996zu} of $M\times F$ (only involving bosons and gravity) in our tensor product extension can be written as
\begin{eqnarray}
  S_b &=& \mathrm{Tr}(f(\triangle D_A/\Lambda)) \nonumber \\
    &=& \frac{1}{\pi^2}(48f_4 \Lambda^4-f_2\Lambda^2\mathrm{Tr}(|G|^2)+\frac{f_0}{4}d\mathrm{Tr}(|G|^4))\int\sqrt{g}d^4x \nonumber \\
   &+&\frac{1}{24\pi^2}(96f_2\Lambda^2-f_0c\mathrm{Tr}(|G|^2))\int R\sqrt{g}d^4x \nonumber \\
   &+& \frac{f_0}{10\pi^2}\int (\frac{11}{6}R^*R^*-3C_{\mu\nu\rho\sigma}C^{\mu\nu\rho\sigma}\mathrm{Tr}(|G|^4))\sqrt{q}d^4x \nonumber \\
   &+&\frac{1}{\pi^2}(-2af_2 \Lambda^2+ef_0\mathrm{Tr}(|G|^4))\int|\varphi|^2\sqrt{g}d^4x \nonumber \\
   &+& \frac{f_0}{2\pi^2}\int a\mathrm{Tr}(|G|^2)|D_\mu \varphi|^2\sqrt{g}d^4x - \frac{f_0}{12\pi^2}\int a\mathrm{Tr}(|G|^2)R|\varphi|^2 \sqrt{g}d^4x \nonumber \\
   &+& \frac{f_0}{2\pi^2}\int b\mathrm{Tr}(|G|^4)|\varphi|^4 \sqrt{g}d^4x \nonumber \\
   &+& \frac{f_0}{2\pi^2}\mathrm{Tr}(|G|^2)\int (g_3^2G_{\mu\nu}^i\bar{G}^{\mu\nu i}+g_2^2F_{\mu\nu}^\alpha\bar{F}^{\mu\nu\alpha} +\frac{5}{3}g_1^2B_{\mu\nu}\bar{B}^{\mu\nu})\sqrt{g}d^4x. \label{S.bi}
\end{eqnarray}
From the assumption that $|G|=1$, we have
\begin{eqnarray}
  S_b &=& \frac{1}{\pi^2}(48f_4 \Lambda^4-f_2\Lambda^2+\frac{f_0}{4}d)\int\sqrt{g}d^4x +\frac{1}{24\pi^2}(96f_2\Lambda^2-f_0c)\int R\sqrt{g}d^4x \nonumber \\
   &+& \frac{f_0}{10\pi^2}\int (\frac{11}{6}R^*R^*-3C_{\mu\nu\rho\sigma}C^{\mu\nu\rho\sigma})\sqrt{q}d^4x
   +\frac{1}{\pi^2}(-2af_2 \Lambda^2+ef_0)\int|\varphi|^2\sqrt{g}d^4x \nonumber \\
   &+& \frac{f_0}{2\pi^2}\int a|D_\mu \varphi|^2\sqrt{g}d^4x - \frac{f_0}{12\pi^2}\int aR|\varphi|^2 \sqrt{g}d^4x + \frac{f_0}{2\pi^2}\int b|\varphi|^4 \sqrt{g}d^4x \nonumber \\
   &+& \frac{f_0}{2\pi^2}\int (g_3^2G_{\mu\nu}^i\bar{G}^{\mu\nu i}+g_2^2F_{\mu\nu}^\alpha\bar{F}^{\mu\nu\alpha} +\frac{5}{3}g_1^2B_{\mu\nu}\bar{B}^{\mu\nu})\sqrt{g}d^4x. \label{S.bi.1}
\end{eqnarray}
That is, the (bosonic and gravity part of) physical action is unchanged by our extension.

The representations of fermions in the original geometry are elements of $\mathcal{H}^{+}=\{\xi \in\mathcal{H}|\gamma\xi=\xi\}$~\cite{Connes:1996gi,Chamseddine:2010ud}. In our extended geometry, $\xi=\eta\otimes\theta\otimes\theta$ ($\eta \in L^2(M,S)$). As discussed above we can denote the factors of $\xi$ restricted to the coordinate fiber space as $\xi'=\eta\otimes\theta(x)$, and rewrite $\xi$ as the tensor products of $\xi'$ and $\theta(1)$.

The fermionic part of the spectral action then reads
\begin{eqnarray} \label{S.f}
& &\frac{1}{2}\sum\langle J \tilde{\eta}\otimes\theta\otimes\theta|\partial\!\!\!/\otimes 1+\gamma_5 \otimes \triangle D_A|\tilde{\eta}\otimes\theta\otimes\theta\rangle \nonumber \\
&=&\frac{1}{2}\sum\langle (J_M\otimes J_F) \tilde{\xi}'|\partial\!\!\!/\otimes 1|\tilde{\xi}'\rangle \nonumber \\
&+& \frac{1}{2}\sum \langle J \tilde{\xi}'|\gamma_5 \otimes \triangle D_A|\tilde{\xi}'\rangle \langle J_F\theta(1)|G|\theta(1)\rangle,
\end{eqnarray}
where $\tilde{\xi}'$ are the Grassmannian analogues of $\xi'$ and $G$ comes from the inner fluctuations of $D$. One can directly check that the factor $\langle J_F \theta|G|\theta\rangle$ of the action is a real scalar coefficient and can be absorbed by the corresponding coupling coefficients.

Thus, the tensor product extension in usual conditions changes neither the bosonic part nor the fermionic part of the physical action. As mentioned, one can always safely make this extension without physical effects.

\section{Quaternion extension}
\label{sec2}

Now we make another extension on the action and the fields in the geometric framework. In particular, we allow for quaternionic results of the action and express the fields in quaternion spaces. Such extended action still has physical meaning and is not changed in usual conditions. Note that there is a trivial quaternion extension, i.e., to map all the complex concepts to those restricted to a slice domain of a complex plane of a quaternion space. We consider a nontrivial, but almost trivial (in usual conditions), quaternion extension. Then we show that when considering the combination of the above two extensions, there can be physical effect, i.e., the emergence of 3 fermion generations.

In this work, we adjust the quaternion extension in two ways, i.e., to adjust the multiplication rules of extended fields and to adjust the imaginary units of the quaternion numbers, to make the extension reflect the physical reality.

\subsection{The Field Adjusted Quaternion Extension}
\label{sec.2.1}

Our field adjusted quaternion extension (quaternion extension for short in this sub-section) on fields and the physical action is as follows:

(i) For fields in the coordinate fiber space of the bundle, we extend the imaginary unit $i$ of plane waves $e^{-ipx}$ (and the corresponding $i$ coupling with these terms in the action) of fields to the imaginary units $I$ in quaternion spaces, and extend the $i$ of $e^{ipx}$ (and the corresponding $i$ coupling with these terms in the action) to the imaginary units $J$ ($J\perp I$) in quaternion spaces, and then label the creation and annihilation operators $\hat{a}, \hat{a}^\dag$ by the corresponding imaginary units as $\hat{a}_I, \hat{a}_J^\dag$. Then, we adjust the multiplication rules of operators $\hat{a}_I, \hat{a}_J^\dag$ by a spin-dependent nonlinear anticommutator involving the extended imaginary units.

(ii) For fields $\theta(1)$ and factors $G$ in the non-coordinate base space of the bundle, we directly extend $\langle J_F \theta|G|\theta\rangle$ to quaternion numbers. In particular, we rewrite the real results of $\langle J_F \theta|G|\theta\rangle$ as
\begin{equation} \label{G.q}
\langle J_F \theta|G|\theta\rangle =r = q+\bar{q}, \quad \mathrm{where}\,\,\, r\in \mathbb{C},\,\,\, q\in \mathbb{H},
\end{equation}
and let $q$, $\bar{q}$ be two independent quaternion numbers (with the relationship $|q+\bar{q}|^2=1$ fulfilled) restricted to the quaternion space spanned by $\{1,I,J,IJ\}$, where $I$ and $J$ come from $\hat{a}_I$ and $\hat{a}_J^\dag$ coupling with $\langle J_F \theta|G|\theta\rangle$. Since  $\theta(1)$ and $G$ are free of local variables and can not expand into plane waves, the multiplication rules of these extended concepts are not adjusted in the way for local fields.

Since the extension on the base space and the fiber should in some sense come from a comprehensive manipulation on the total space of the bundle, it is a natural condition that $q$ and $\bar{q}$ are restricted to the quaternion space $\mathbb{H}_{IJ}=\mathbb{R}\oplus I\mathbb{R} \oplus J\mathbb{R} \oplus IJ\mathbb{R}$, where $I,J$ come from the corresponding $\hat{a}_I, \hat{a}_J^\dag$. When our quaternion extension is generalized to higher dimensional number extensions, i.e., to let $I$ and $J$ of $\hat{a}_I$ and $\hat{a}_J^\dag$ be the imaginary units of higher dimensional numbers, the restriction of $q$ and $\bar{q}$ ensures that the result of this work (i.e., the number of fermion generations) keeps unchanged. In other words, our extension is actually a ``$2^n$-dimensional'' ($n>2$) extension and the extended imaginary units play an essential role. For convenience of discussion, we mainly focus on the quaternion extension in the minimal nontrivial case, while the result is not changed.

(iii) According to the above extensions, the results of traces and forms $\langle \rangle$ (where we simplify forms of $\langle \rangle$ applied to operators and fields directly as $\langle \rangle$) are allowed to be quaternion numbers. The extended quaternion results of traces are shown later to be equal to ordinary complex numbers or complex matrices, and thus the physical meaning of the action is not changed by our extension. The complex matrices as results of traces come from $\langle J_F \theta(1)|G|\theta(1)\rangle$, and 3 fermion generations emerge from these matrices.

Fermion fields $\xi'$ can expand into plane waves $u^s(p)e^{-ipx}$ and $v^s(p)e^{ipx}$ with operators $\hat{a}_p^s$ and $\hat{b}_p^{s\dag}$ as the corresponding coefficients. As mentioned above, we extend $i$ of the above plane waves to the imaginary units of quaternion spaces. Then we map $\hat{a}$ and $\hat{a}^{\dag}$ to $\hat{a}_I$ and $\hat{a}_J^\dag$ (while the labels $s$ and $p$ are ignored), where $I, J \in \mathbb{S}_a =\{q \in \mathbb{H}_a| q^2=-1\}$ with $I$$\perp$$J$. Similarly, $\hat{b}$ and $\hat{b}^{\dag}$ are mapped to $\hat{b}_I$ and $\hat{b}_J^\dag$, where $I, J \in \mathbb{S}_b =\{q\in\mathbb{H}_b |q^2=-1\}$ with $I$$\perp$$J$. The extended operators are normalized slice preserving quaternionic functions. This ensures that the spectral action principle~\cite{Connes:1988ym, Chamseddine:1996zu} still works. Furthermore, for any $K_a \in \mathbb{S}_a$ and $K_b \in \mathbb{S}_b$, we let $K_a \perp K_b$.

Now fields $\xi'$ (or $(J_M\otimes J_F)\xi'$) are determined by operators $(\hat{a}_I)_{\bf p}^s$ and $(\hat{b}_J^{\dag})_{\bf p}^s$ (or $(\hat{a}_J^\dag)_{\bf p}^s$ and $ (\hat{b}_I)_{\bf p}^s$), i.e., the quaternion functions on slices $(L_I)_a \subset \mathbb{H}_a$ and $(L_J)_b \subset \mathbb{H}_b$ (or $(L_J)_a \subset \mathbb{H}_a$ and $(L_I)_b \subset \mathbb{H}_b$). One can always define quaternion spaces $\mathbb{H}_1, \mathbb{H}_2$ with any $I_1 \subset \mathbb{H}_1$ and $I_2 \subset \mathbb{H}_2$ fulfilling $I_1 \perp I_2$, such that $(L_I)_a, (L_J)_b \subset \mathbb{H}_1$ and $(L_J)_a, (L_I)_b \subset \mathbb{H}_2$. In other words, $\xi'$ (or $(J_M\otimes J_F) \xi'$) are fields on quaternion spaces $\mathbb{H}_1$ (or $\mathbb{H}_2$), and $(J_M\otimes J_F)$ maps the fields on $\mathbb{H}_1$ and those on $\mathbb{H}_2$ to each other.

Although the representations of operators are extended to be quaternionic, their multiplication rules are not directly identical to those of the quaternion numbers to keep the extension ``almost'' trivial. The multiplications of operators are given by actions of operators from right to left sequently on states of fields in the common way. Specifically, the multiplications of creation and annihilation operators with label $\alpha$, for both fermions and bosons, are defined as
\begin{eqnarray}
&&(\hat{a}^{s\dag}_{\phantom{\dag}J})_\alpha |\xi'\rangle =\frac{1}{\sqrt{N+1}} \xi'_\alpha \otimes_{(-1)^{2s}} |\xi'\rangle, \label{a.s} \\
&&(\hat{a}_I^{s})_\alpha |\xi' \rangle =\frac{1}{\sqrt{N}} \xi'_\alpha \oslash_{(-1)^{2s}} |\xi'\rangle, \label{a.s.+}
\end{eqnarray}
where $s$ is the spin of $\xi'$, and the insertion $\otimes_\pm$ and the deletion operator $\oslash_\pm$ are defined as
\begin{eqnarray}
&& \xi'_\alpha \otimes_{\pm} 1 = \xi'_\alpha, \quad \xi'_\alpha \oslash_{\pm} 1 = 0, \label{o.01} \\
&&\xi_\alpha' \otimes_\pm (\xi'_\beta \otimes \xi')=\xi'_\alpha\otimes\xi'_\beta\otimes \xi' \pm \xi'_\beta \otimes (\xi'_\alpha \otimes_\pm \xi'), \label{o+} \\
&& \xi_\alpha' \oslash_\pm (\xi'_\beta \otimes \xi')= \delta_{\alpha \beta }\xi' \pm \xi'_\beta \otimes (\xi'_\alpha \oslash_\pm \xi'). \label{o-}
\end{eqnarray}
These multiplication rules are irrelevant to the labels $I$ and $J$, and return to the complex operator multiplication rules. For normalized slice preserving functions, one can always define such multiplication rules. However, the results of $\xi'_\alpha \otimes_\pm 0$ and $\xi'_\alpha \oslash_\pm 0$, which come from $\hat{c}^\dag_J d_I$ and $\hat{c}_I d_I $ ($\hat{c}, \hat{d} \in \{\hat{a}, \hat{b}\}$), are not defined by these multiplication rules. Our quaternion extension can only manifest itself on multiplication rules involving $\hat{c}^\dag_J \hat{d}_I$ and $\hat{c}_I \hat{d}_I$. In canonical quantization process, these undefined multiplications are adjusted by quantum (anti-)commutation rules. In our scheme, we adjust the undefined rules of both fermions and bosons by a spin-dependent ``anticommutator'' $\{\cdot, \cdot \}_s$ defined as
\begin{equation} \label{anti.n}
\{A^t, B^r \}_s = A^t\circ B^r -(-)^{t+r} B^r \circ A^t,
\end{equation}
which is mainly determined by quaternion extensions on operators. Here $A$ and $B$ are locally defined quaternion extended fields and $t$ and $r$ are the corresponding spin labels.

For fermions, this anti-commutator gives the anticommutative rules of creation and annihilation operators. The computation of the anticommutator $\{\hat{c}^t_I(\zeta), \hat{d}^r_J(\eta)\}_s$ ($\hat{c}, \hat{d} \in \{ \hat{a}, \hat{a}^\dag, \hat{b}, \hat{b}^\dag\}$, $\zeta, \eta \in M$, $t,r$ are spin labels) is defined as follows: for $I, J \in \mathbb{S}_a$ or $I, J \in \mathbb{S}_b$, and $\hat{c}=\hat{d}$ or $\eta=\zeta, t=r$,
\begin{equation} \label{anti.f.1}
\{\hat{c}^t_I(\zeta), \hat{d}^r_J(\eta)\}_s = \{ \hat{c}^t_I(\zeta), \hat{d}^r_J(\eta) \} = (2\pi)^3 e^{\sum_{I,J}\{I,J\}},
\end{equation}
and for other cases
\begin{equation} \label{anti.f.2}
\{\hat{c}^t_I(\zeta), \hat{d}^r_J(\eta)\}_s = \{ \hat{c}^t_I(\zeta), \hat{d}^r_J(\eta) \} = (2\pi)^3 \sum_{I,J}\{I,J\}.
\end{equation}
The factor $(2\pi)^3$ is given for the normalization of fields. Since the directions of $I$ and $J$ in $\mathbb{H}$ are not determined by our quaternion extension, which only restricts relationships of $I$ and $J$, there can be infinite directions of possible $I$ and $J$ to be summated. Counting the infinite numbers of $I$ and $J$, one can directly check that (\ref{anti.f.1}) and (\ref{anti.f.2}) give the canonical quantum anticommutation rules of creation and annihilation operators of fermions. Then, the fermions in the coordinate fiber space under our quaternion extension are equivalent to those of the standard model (without the assumption of {\it ad hoc} 3 generations).

The creation and annihilation operators (which are also denoted as $\hat{a}$ and $\hat{a}^\dag$ in this part) of bosons are similarly extended: $\hat{a} \mapsto \hat{a}_I$, $\hat{a}^\dag \mapsto \hat{a}^\dag_{\phantom{i}J}$, where $I,J \in \mathbb{S}_a=\{q\in \mathbb{H}| q^2=-1 \}_a$ with $I$$\perp$$J$. The multiplications of these operators on states, as defined above, are
\begin{eqnarray}
&& (\hat{a}^\dag_{\phantom{i}J})_\alpha | \phi \rangle = \frac{1}{\sqrt{N+1}} \phi_\alpha \otimes_+ |\phi \rangle, \label{a.b} \\
&& (\hat{a}_I)_\alpha | \phi \rangle = \frac{1}{\sqrt{N}} \phi_\alpha \oslash_+ | \phi \rangle. \label{a.b.+}
\end{eqnarray}
The spin-dependent nonlinear anticommutator of bosonic operators $\hat{c}_I(\zeta)$ and $\hat{d}_{J}(\eta)$ ($\hat{c}, \hat{d} \in \{\hat{a}, \hat{a}^\dag\}$) is defined as: if $\hat{c}=\hat{d}$ or $\eta=\zeta$,
\begin{equation}\label{anti.b.1}
\{\hat{c}_I(\zeta), \hat{d}_{J}(\eta)\}_s = [\hat{c}_I(\zeta), \hat{d}_J(\eta)] = (2\pi)^3 e^{\sum_{I,J}\{I,J\}},
\end{equation}
and in other cases
\begin{equation}\label{anti.b.2}
\{\hat{c}_I(\zeta), \hat{d}_{J}(\eta)\}_s = [\hat{c}_I(\zeta), \hat{d}_J(\eta)] = (2\pi)^3\sum_{I,J}\{I,J\}.
\end{equation}
Just like the fermion case, one can directly check that these are the canonical quantum  commutation rules of bosons. Then, the quaternion extended bosons in the coordinate fiber space are equivalent to those of the standard model.

As shown above, our quaternion extension on the standard model, when restricted to the coordinate fiber, does not have physical effect. Now we combine the quaternion and tensor product extensions. That is, we make quaternion extension on the global factors $\langle J_F \theta(1)|G|\theta(1)\rangle$ of the action coming from the non-coordinate structure under the tensor product extension. In usual conditions, these factors are global scalar coefficients on the action. As mentioned above, we extend these coefficients to quaternion numbers $Q\in \mathbb{H}_{IJ}$ ($|Q|=1$). Without extra assumptions, $Q$ are not restricted by other rules. Thus, $Q$ are nontrivial quaternion numbers, and can have different constant values when coupling with different particles. We denote the factor coupling with Yukawa terms of quarks by $Q_1$, and that of leptons by $Q_2$.

Let the representation of quark in the coordinate fiber space be function in $\mathbb{H}_1$ after the quaternion extension. With given $(I,J)\in \mathbb{H}_1$, this extended representation can be uniquely labelled by the complementary space (which is denoted as $L_{IJ}$ with $IJ \perp I,J$) of the plane generated by $I,J$ in the quaternion space. Since the directions of $I,J$ in $\mathbb{H}_1$ are not definitely determined, there are extended representations of quarks labelled by $L_I$ and $L_J$ and these representations are equal to those labelled by $L_{IJ}$. Thus the extended quarks are splitted into
\begin{equation}\label{q.IJ}
  \xi_q' = (\xi_q')_{L_{IJ}} \oplus (\xi_q')_{L_{I}} \oplus (\xi_q')_{L_{J}}.
\end{equation}
The extended leptons in $\mathbb{H}_2$ can be similarly splitted into
\begin{equation}\label{l.IJ}
  \xi_l' = (\xi_l')_{L_{IJ}} \oplus (\xi_l')_{L_{I}} \oplus (\xi_l')_{L_{J}}.
\end{equation}
Without the tensor product extension, since the imaginary units in $(L_{IJ}, L_{I}, L_{J})_q$ and $(L_{IJ}, L_{I}, L_{J})_l$ do not manifest the ordinary multiplication rules of quaternion numbers, (\ref{q.IJ}) and (\ref{l.IJ}) do not give explicit contribution to the action.

$Q_{1}$ and $Q_{2}$ multiply on the action by the ordinary quaternion multiplication. This quaternion multiplication only manifests itself by mapping the vectors $(L_{IJ},$ $ L_{I},$ $ L_{J})_q$ and $(L_{IJ}, L_{I}, L_{J})_l$ to some $(L_{I'J'}, L_{I'}, L_{J'})_q$ and $(L_{I'J'}, L_{I'}, L_{J'})_l$. With the condition $|Q|=1$ (without this condition, one just need a reparameterization of $\mathbb{H}_1, \mathbb{H}_2$, and corresponding coupling coefficients in the action), one can safely consider $(L_{IJ}, L_{I}, L_{J})_q$ and $(L_{IJ}, L_{I}, L_{J})_q$ as complex vectors multiplied on $\xi_q'$ and $\xi_l'$, and consider $Q_1$ and $Q_2$ as the complex rotation matrices on those vectors.

The gauge bosons and the Higgs are free of the factors $Q$ (since only terms with $|Q|^{2n}$ give contribution to the bosonic part of the spectral action~\cite{nc.rev}). Thus, the vectors $(L_{IJ}, L_{I}, L_{J})$ of bosons give trivial contributions on the action, and our extensions do not change the bosonic part of the action.

Only the Higgs and fermion coupling terms of Lagrangian from (\ref{S.f}) has explicit effect from the combined extensions. We write this part of Lagrangian as
\begin{equation} \label{fTf}
\mathcal{L}_{Hf} = - \bar{f}_l T(K_\nu,K_e, \varphi)Q_2 f_l - \bar{f}_q T(K_u, K_d, \varphi) Q_1 f_q,
\end{equation}
where
\begin{eqnarray}
  f_l &=& (|\!\!\uparrow\rangle\otimes {\bf 1}^0, |\!\!\downarrow\rangle\otimes {\bf 1}^0)^T_{L_{I}}\oplus (|\!\!\uparrow\rangle\otimes {\bf 1}^0, |\!\!\downarrow\rangle\otimes {\bf 1}^0)^T_{L_{IJ}} \oplus (|\!\!\uparrow\rangle\otimes {\bf 1}^0, |\!\!\downarrow\rangle\otimes {\bf 1}^0)^T_{L_{J}} \nonumber \\
   &=& ((\nu_e, e)^T,(\nu_\mu, \mu)^T,(\nu_\tau, \tau)^T)^T, \label{lep.3}\\
  f_q &=& (|\!\!\uparrow\rangle\otimes {\bf 3}^0, |\!\!\downarrow\rangle\otimes {\bf 3}^0)^T_{L_{I}}\oplus (|\!\!\uparrow\rangle\otimes {\bf 3}^0, |\!\!\downarrow\rangle\otimes {\bf 3}^0)^T_{L_{IJ}} \oplus (|\!\!\uparrow\rangle\otimes {\bf 3}^0, |\!\!\downarrow\rangle\otimes {\bf 3}^0)^T_{L_{J}} \nonumber \\
   &=& ((u, d)^T,(c, s)^T,(t, b)^T)^T, \label{qua.3}
\end{eqnarray}
and
\begin{equation} \label{T.k12}
T(K_1,K_2,\varphi)=\left(\begin{array}{cccc}
0&0&K_1^*\varphi_1&K_1^*\varphi_2 \\
0&0&-K_2^*\bar{\varphi}_2&K_2^*\bar{\varphi}_1 \\
K_1\bar{\varphi}_1&-K_2\varphi_2 &0 & 0 \\
K_1\bar{\varphi}_2&K_2\varphi_1 &0 & 0
\end{array} \right),
\end{equation}
is deduced from operator $A^{(0,1)}_0$ in the same way of~\cite{nc.rev,Chamseddine:2006ep,Connes:2006qv,Chamseddine:2010ud,Connes:1996gi}, with coefficients $\varphi, K_\nu, K_e, K_u, K_d$ formally the same as those in~\cite{nc.rev}:
\begin{eqnarray}
& &\varphi_1 = \phi_1 + \frac{2M}{g}, \quad \varphi_2 = \phi_2, \label{hig.12} \\
& & (K_\nu)_{\mu\rho}= \frac{g}{2M}m_\nu^{\mu}\delta_\mu^\rho, \quad
(K_e)_{\mu\rho} = \frac{g}{2M}m_e^\mu C_{\mu\lambda}^{\mathrm{lep}} \delta^\kappa_\lambda (C^{\mathrm{lep}})_{\kappa\rho}^\dagger, \label{C.lep} \\
& & (K_u)_{\mu\rho} = \frac{g}{2M}m_u^\mu\delta^\rho_\mu, \quad
(K_d)_{\mu\rho} = \frac{g}{2M}m_d^\lambda C_{\mu\lambda} \delta_\lambda^\kappa C_{\kappa\rho}^\dagger. \label{C.qua}
\end{eqnarray}
We suppose $Q_1$ and $Q_2$, as 3-dimensional rotation matrices (with $|Q_1|^2=1$ and $|Q_2|^2=1$), are diagonalizable, i.e.,
\begin{eqnarray}
Q_1 &=& C_{q1} \left(
\begin{array}{ccc}
\alpha_1 & 0 & 0 \\
0 & \alpha_2 & 0 \\
0 & 0 & \alpha_3
\end{array}
\right)C_{q2}, \label{Mq.0} \\
Q_2 &=& C_{l1}\left(
\begin{array}{ccc}
\beta_1 & 0 & 0 \\
0 & \beta_2 & 0 \\
0 & 0 & \beta_3
\end{array}
\right)C_{l2}, \label{Ml.0}
\end{eqnarray}
where $C_l, C_q$ are invertible rotation matrices. Thus the masses of quarks and leptons are
\begin{eqnarray}
  \left(
  \begin{matrix}
    m_u & 0 & 0 \\
    0 & m_c & 0 \\
    0 & 0 & m_t
  \end{matrix}
  \right) &=& \left(
\begin{array}{ccc}
\frac{g}{2M}\Upsilon_u \bar{\phi}_1 \alpha_1 & 0 & 0 \\
0 &\frac{g}{2M} \Upsilon_u \bar{\phi}_1 \alpha_2 & 0 \\
0 & 0 &\frac{g}{2M} \Upsilon_u \bar{\phi}_1 \alpha_3
\end{array}\right), \\
  \left(
  \begin{matrix}
    m_d & 0 & 0 \\
    0 & m_s & 0 \\
    0 & 0 & m_b
  \end{matrix}
  \right) &=& \left(
\begin{array}{ccc}
\frac{g}{2M}\Upsilon_d \phi_1 \alpha_1 & 0 & 0 \\
0 & \frac{g}{2M}\Upsilon_d \phi_1 \alpha_2 & 0 \\
0 & 0 & \frac{g}{2M}\Upsilon_d \phi_1 \alpha_3
\end{array}
\right), \\
  \left(
  \begin{matrix}
    m_{\nu_e} & 0 & 0 \\
    0 & m_{\nu_\mu} & 0 \\
    0 & 0 & m_{\nu_\tau}
  \end{matrix}
  \right) &=& \left(
\begin{array}{ccc}
\frac{g}{2M}\Upsilon_\nu \bar{\phi}_1 \beta_1 & 0 & 0 \\
0 & \frac{g}{2M}\Upsilon_\nu \bar{\phi}_1 \beta_2 & 0 \\
0 & 0 & \frac{g}{2M}\Upsilon_\nu \bar{\phi}_1 \beta_3
\end{array}
\right), \label{nu.m} \\
\left(
  \begin{matrix}
    m_{e} & 0 & 0 \\
    0 & m_{\mu} & 0 \\
    0 & 0 & m_{\tau}
  \end{matrix}
  \right)  &=& \left(
\begin{array}{ccc}
\frac{g}{2M}\Upsilon_e \phi_1 \beta_1 & 0 & 0 \\
0 & \frac{g}{2M}\Upsilon_e \phi_1 \beta_2 & 0 \\
0 & 0 & \frac{g}{2M}\Upsilon_e \phi_1 \beta_3
\end{array}
\right).
\end{eqnarray}
The flavor mixing of leptons and that of quarks are characterized by $C_{qi}$, $C_{li}$ ($i \in \{1,2\}$), together with the matrices $C$ and $C^{\mathrm{lep}}$ in (\ref{C.qua}) and (\ref{C.lep}) in the classical noncommutative geometry as the mixing matrices in 1 generation.

Thus we construct the standard model with $3$ fermion generations as the output of our extensions, but not as an {\it ad hoc} input parameter. The 3 generation of fermions are formally the same with physical elementary fermions with the right gauge coupling charges. The CKM and PMNS matrix parameters are hidden in matrices reduced from $Q_2$ and $Q_1$.

We mention that the nonlinear structure in the quaternion extension is highly related to the quantization process and plays an essential role in the fermion generation emergence. In other words, features of elementary particles are connected with the quantization process by the nonlinear anticommutator in our work.

\subsection{Imaginary Unit Adjusted Quaternion Extension}
\label{sec.2.2}

There are different ways to make the quaternion extended framework to reflect physical reality. As discussed in this sub-section, one can directly adjust the imaginary units of quaternion numbers in the quaternion extension. The imaginary units are invariant in common sense. However, in the discussion after (\ref{G.q}), quaternion numbers are assumed to be able to come from the independence between $q$ and $\bar{q}$. One can consider this independence as an analogue of the independence between physical operators $\hat{a}$ and $\hat{a}^\dag$. Since the physical operators are affected by the physical nature, the quaternion space can be also considered to be possibly affected by the physical framework. In this sub-section, the extension on the fields and the action restricted to the non-coordinate space of the bundle is the same as that in Sub-Section~\ref{sec.2.1}, and the independence between $q$ and $\bar{q}$ keeps unchanged. Thus, the imaginary units can be correspondingly adjusted.

Our imaginary unit adjusted quaternion extension is as follows:

(i) We extend the fields in the coordinate space of the bundle to quaternionic fields. Then, we adjust the imaginary units of the quaternion space by investigating the Higgs mechanism in this extended geometry.

In this extension, any term of the action falls into a specific complex slice of the quaternion space. We assume that the coupling strength of one term of the action is proportional to the square of the norm of the adjusted imaginary unit of the corresponding complex slice. We mention that a term of the action can be on the real axis of the quaternion space. In this case, one can not tell which complex slice this term actually falls into. However, the coupling fields in this real term should be in one complex slice, and we assume that the coupling strength of this term is proportional to the square of the norm of the adjusted imaginary unit of that complex slice.

(ii) Our quaternion extension on fields $\theta(1)$ and factors $G$ in the non-coordinate space of the bundle is the same as that in Sub-Section~\ref{sec.2.1}.

Now we investigate the Higgs mechanism in the quaternion extension. One typical potential of the Higgs mechanism can be written as~\cite{qft}
\begin{equation}\label{poten.Hig.0}
  V(\phi)=-\mu^2\phi^*\phi+\frac{\lambda}{2}(\phi^*\phi)^2,
\end{equation}
which is invariant under the local U$(1)$ transformation on $\phi$. Here $\phi$ is a scalar field. In the ordinary standard model, $\mu$ and $\lambda$ are real coupling parameters. Once the system falls into a specific minimum of the potential $V(\phi)$ at a non-zero point of $\phi$, the U$(1)$ global symmetry of the system will be spontaneously broken. If the above minimum is stable, $\phi$ will acquire a non-zero vacuum expectation value and particles in the standard model will acquire masses through the Higgs mechanism. To ensure the stability of the minimum of $V(\phi)$, $\lambda$ should be restricted to positive real numbers. For convenience of discussion, we rewrite the potential $V(\phi)$ as
\begin{equation}\label{poten.Hig.i}
  V(\phi)=\omega^2\phi^*\phi+\frac{\kappa^2}{2}(\phi^*\phi)^2,
\end{equation}
where $\omega=I \mu$ (here $I$ is the imaginary unit) and $\kappa=\sqrt{\lambda}\in \mathbb{R}$. In other words, $\kappa$ as a vector on the complex plane is perpendicular to $\omega$ which is on the imaginary axis.

In our quaternion extended framework, there are three imaginary axes. Then, both $\omega$ and $\kappa$ can be on imaginary axes with the condition $\omega \perp \kappa$ fulfilled. For instance, one can let $\omega=J \mu$ and $\kappa= I \sqrt{\lambda}$, where $I$, $J$ and $K$ are the imaginary units of the quaternion space. In this case, the potential $V(\phi)$ is rewritten as
\begin{equation}\label{poten.Hig.ij.1}
  V(\phi)=-\mu^2 \phi^*\phi - \frac{\lambda}{2} (\phi^* \phi)^2,
\end{equation}
which has no stationary point. Consequently, the Higgs mechanism can not emerge in this extended framework. We mention that the possibility of defining $\kappa$ as the form of $I \sqrt{\lambda}$ is directly given by our quaternion extension. To ensure that the Higgs mechanism can emerge, there should be some extra structure in the quaternion extension.

In this work, the structure of the extended framework to ensure the emergence of the Higgs mechanism is the adjustment on the imaginary units. Now we consider the imaginary units $I$, $J$, and $K$ of the quaternion space as analogues of fields, which can be acted and characterized by operators. Then, we adjust the imaginary units as follows:
\begin{equation}\label{IJK}
  I \mapsto I' \equiv J' K', \quad J \mapsto J^\prime=Ie^{-\theta_1}, \quad \text{and}\,\,\,\, K\mapsto K^\prime= I e^{\theta_2},
\end{equation}
where $\theta_1$ and $\theta_2$ are positive real numbers. We mention that there is no composition of adjustments on the imaginary units. That is, the adjustment $I \mapsto I'$ is not made in $J\mapsto J'=Ie^{-\theta_1}$. Thus, $I^{\prime 2}=(J^\prime K')^2=I^4 \delta=\delta$, with $\delta=e^{2(\theta_2-\theta_1)}$, and $J^2= I^2 e^{-2\theta_1}=-e^{-2\theta_1}$. Then, the potential $V(\phi)$ in (\ref{poten.Hig.i}) reads
\begin{equation}\label{poten.Hig.ii}
  V(\phi)=-\mu^{\prime 2} \phi^*\phi+\frac{\lambda '}{2}(\phi^*\phi)^2,
\end{equation}
where $\mu^\prime = e^{-2\theta_1}\mu$ and $\lambda '=\delta \lambda$. After a reparameterization, $V(\phi)$ in (\ref{poten.Hig.ij.1}) is equivalent with that in (\ref{poten.Hig.0}), and the corresponding particles can acquire masses through the Higgs mechanism.

Our adjustment on the imaginary units not only ensures that the Higgs mechanism can emerge in the quaternion extended framework, but also plays an essential role in establishing the relationships among norms of the adjusted imaginary units. 

One relationship among norms of the adjusted imaginary units can be directly read from (\ref{IJK}):
\begin{equation}\label{IJK.<}
  |J'|< |I'|< |K'|, \quad \text{with} \,\,\, \frac{|I'|}{|J'|}=e^{\theta_2}, \,\,\, \frac{|K|'}{|I'|}=e^{\theta_1}.
\end{equation}

Another relationship among norms of the adjusted imaginary units comes from an investigation into the complex slices of those imaginary units. For a complex slice $L_{\vartheta}$ (with $\vartheta\in \{I' ,\,\,J' ,\,\,K' \}$) of the quaternion space, the total quantity of points in this slice can be characterized by the area $S_\vartheta$ of the slice. This area is obviously an infinite number. The cardinality of $S_\vartheta$ can be characterized by $|\vartheta|^2 \aleph$, where $\aleph$ denotes the cardinality of the area of the complex slice of the ordinary quaternion space. Mathematically, $|\vartheta|^2 \aleph$ is equal to $\aleph$. Nevertheless, to reflect physical nature, there can be a finite cut-off of $\aleph$, and then $|\vartheta|^2 \aleph$ is different from $\aleph$. Thus, the scale of $S_\vartheta$ splits into two parts, i.e., the cardinality $\aleph$ corresponding to the ordinary quaternion space and the scale corresponding to the adjusted imaginary units. We denote the scale corresponding to the adjusted imaginary units as $s_\vartheta$. Since $\aleph$ is not affected by the adjustment on the imaginary units, it is natural to consider $\aleph$ keeps unchanged in different complex slices. Then, the differences among the areas of complex slices are manifested by the scales $s_\vartheta$ of those slices.

The complex slices $L_{I'}$, $L_{J'}$ and $L_{K'}$ of the quaternion space of adjusted imaginary units can be investigated from two aspects. The first aspect is to consider the above complex slices as a set whose elements are the three complex slices. The second aspect comes from the adjustment on the imaginary units. Since the adjusted imaginary units are all proportional to $I$ (or the norm of $I$), the corresponding complex slices can be considered to form a new complex slice, whose imaginary unit is the summation of $I'$, $J'$, and $K'$.  The total quantity of points in the adjusted complex slices should be invariant under investigations from the above two aspects. In other words, $s_{\mathrm{sum}}$ (which denotes the scale $s_\vartheta$ of the above three complex slices) calculated from the two aspects are equal to each other.

In the first aspect, $s_{\mathrm{sum}}$ reads,
\begin{equation}\label{s.ijk}
  s_{\mathrm{sum}}= \frac{1}{2}(|I'|^2+|J'|^2+|K'|^2).
\end{equation}
We mention that the adjustment of (\ref{IJK}) is based on the even permutation of $I'$, $J'$, and $K'$. To exclude the contribution of odd permutation of the imaginary units to $s_{\mathrm{sum}}$, there is a coefficient $\frac{1}{2}$ in (\ref{s.ijk}).

In the seconde aspect, we have
\begin{equation}\label{s.iii}
  s_{\mathrm{sum}}=\frac{1}{3}(|I'|+|J'|+|K'|)^2.
\end{equation}
Here $I'$, $J'$, and $K'$ play equal roles in the formula $(|I'|+|J'|+|K'|)^2$. To select the specific imaginary unit $I'$ for the adjustment, there is a coefficient $\frac{1}{3}$ in (\ref{s.iii}). Since $s_{\mathrm{sum}}$ calculated from the two aspects are equal to each other, we have
\begin{equation}\label{koide.ijk}
  |I'|^2+|J'|^2+|K'|^2 = \frac{2}{3} (|I'|+|J'|+|K'|)^2.
\end{equation}

In this quaternion extension, an ordinary field $\xi$ is extended to quaternionic field. That is, the field $\xi$ splits into projections $(\xi_{L_{I'}},\xi_{L_{J'}},\xi_{L_{K'}})$ on the three complex slices $L_{I'}$, $L_{J'}$, and $L_{K'}$ of the quaternion space. Because of the adjustment (\ref{IJK}) on the imaginary units, the projections of an extended field on the three complex slices equivalently fall into one complex slice $L_{(e^{\theta_2} + e^{-\theta_1})I}$, which can be considered as $L_{I}$ after a reparameterization. Without rotations on $(L_{I'}, L_{J'}, L_{K'})$, the projections of one extended field on the three complex slices not only fall into one complex slice, but also contribute to the action as a whole. Then, the extended fields are equivalent with those in the ordinary standard model represented by complex fields. Moreover, the adjustment on the imaginary units ensures that terms of the action fall into one complex slice. As mentioned above, coupling coefficients of those terms of the action are proportional to the square of the norm of the imaginary unit of that complex slice. Without rotations on $(L_{I'}, L_{J'}, L_{K'})$, the square of the norm of that imaginary unit can be absorbed by coupling coefficients after a reparameterization. Consequently, the quaternion extension does not manifest itself without being combined with other structures (e.g., the tensor product extension) which can generate rotations on $(L_{I'}, L_{J'}, L_{K'})$.


Now we combine the tensor product extension with the quaternion extension. In other words, we make the quaternion extension on the action restricted to the coordinate space and the action $\langle J_F \theta(1)|G|\theta(1)\rangle$ restricted to the non-coordinate space. As discussed in the above sub-section, this combination multiplies the Yukawa coupling terms in the coordinate space by quaternion global factors $Q=\langle J_F \theta(1)|G|\theta(1)\rangle$. The imaginary units of the quaternion space of the extended geometry restricted to the non-coordinate space are not adjusted. Thus, the quaternion factors $Q$ multiply on the action restricted to the coordinate fiber space by the ordinary quaternion multiplication, which only manifests itself by some rotation on the vector $f=(f_{L_{I'}}, f_{L_{J'}}, f_{L_{K'}})$. Under that rotation, the projections of fermions on the three complex slices are respectively and independently written in the action. That is, the fermions split into three generations. As in Sub-Section~\ref{sec.2.1}, we denote the factor $Q$ coupling with Yukawa terms of quarks by $Q_1$, and that of leptons by $Q_2$. The corresponding Yukawa coupling terms of the Lagrangian read,
\begin{equation} \label{fTf.1}
\mathcal{L}_{Hf} = - \bar{f}_l T(K_\nu,K_e, \varphi)Q_2 f_l - \bar{f}_q T(K_u, K_d, \varphi) Q_1 f_q,
\end{equation}
with
\begin{eqnarray}
  f_l &=& (|\!\!\uparrow\rangle\otimes {\bf 1}^0, |\!\!\downarrow\rangle\otimes {\bf 1}^0)^T_{L_{J'}}\oplus (|\!\!\uparrow\rangle\otimes {\bf 1}^0, |\!\!\downarrow\rangle\otimes {\bf 1}^0)^T_{L_{I'}} \oplus (|\!\!\uparrow\rangle\otimes {\bf 1}^0, |\!\!\downarrow\rangle\otimes {\bf 1}^0)^T_{L_{K'}} \nonumber \\
   &=& ((\nu_e, e)^T,(\nu_\mu, \mu)^T,(\nu_\tau, \tau)^T)^T, \label{lep.3}\\
  f_q &=& (|\!\!\uparrow\rangle\otimes {\bf 3}^0, |\!\!\downarrow\rangle\otimes {\bf 3}^0)^T_{L_{J'}}\oplus (|\!\!\uparrow\rangle\otimes {\bf 3}^0, |\!\!\downarrow\rangle\otimes {\bf 3}^0)^T_{L_{I'}} \oplus (|\!\!\uparrow\rangle\otimes {\bf 3}^0, |\!\!\downarrow\rangle\otimes {\bf 3}^0)^T_{L_{K'}} \nonumber \\
   &=& ((u, d)^T,(c, s)^T,(t, b)^T)^T, \label{qua.3}
\end{eqnarray}
where the subscripts $L_{\vartheta}$ (with $\vartheta \in \{I',\,\, J',\,\, K' \}$) denote the projections of fermions on the corresponding complex slices. The matrix $T(K_1,K_2, \varphi)$ (with $K_1 \in \{K_\nu,\,\, K_u\}$, $K_2 \in \{K_e,\,\, K_d\}$, and $\varphi= (\varphi_1,\,\, \varphi_2)^T$) and the corresponding variables $K_1$, $K_2$, and $\varphi_2$ in (\ref{fTf.1}) are defined in the same formulae as those in (\ref{T.k12}), (\ref{hig.12}), (\ref{C.lep}) and (\ref{C.qua}), while the variable $\varphi_1$ in $T(K_1,K_2, \varphi)$ is defined as follows: when $\varphi_1$ in (\ref{fTf.1}) is coupling with $f_{L_{\vartheta }}$,
\begin{equation}\label{phi1.1}
  \varphi_1=\phi_1 |\vartheta |^2 + \frac{2M}{g}.
\end{equation}
The factor $|\vartheta|^2$ in (\ref{phi1.1}) comes from our assumption that, for a projection of some Yukawa coupling term on a complex slice $L_{\vartheta}$, the coupling coefficient is proportional to $|\vartheta|^2$. This assumption means that all the points on a complex slice give equal contributions to the Yukawa coupling terms projected on this complex slice.

We mention that the adjusted imaginary units of the extended framework only manifest themselves by the factors $|\vartheta|^2$ of the Yukawa coupling coefficients. Thus, the vectors $(f_{L_{J'}},\,\, f_{L_{I'}},\,\, f_{L_{K'}})$ can be equivalently considered as complex vectors. Moreover, the quaternion numbers $Q_1$ and $Q_2$ only manifest themselves by rotating vectors $(f_{L_{J'}},\,\, f_{L_{I'}},\,\, f_{L_{K'}})$ to some other vectors $(f_{L_{J'}}' ,\,\, f_{L_{I'}}' ,\,\, f_{L_{K'}}')$, and then one can also consider $Q_1$ and $Q_2$ as complex 3-dimensional rotation matrices (with $|Q_i|^2=1$, $i\in \{1,\,\, 2\}$) in the extended action. In others words, the extended action can be equivalently represented by ordinary complex fields, with fermions falling into three generations. As in the above sub-section, we suppose that $Q_1$ and $Q_2$ are diagonalizable, and rewrite them as
\begin{eqnarray}
Q_1 &=& C_{q1} \left(
\begin{array}{ccc}
\alpha_1 & 0 & 0 \\
0 & \alpha_2 & 0 \\
0 & 0 & \alpha_3
\end{array}
\right)C_{q2}, \label{Mq.01} \\
Q_2 &=& C_{l1}\left(
\begin{array}{ccc}
\beta_1 & 0 & 0 \\
0 & \beta_2 & 0 \\
0 & 0 & \beta_3
\end{array}
\right)C_{l2}, \label{Ml.01}
\end{eqnarray}
where $C_l, C_q$ are invertible rotation matrices. Thus, the masses of quarks and leptons are

\begin{eqnarray}
 & & \left(
  \begin{matrix}
    m_u & 0 & 0 \\
    0 & m_c & 0 \\
    0 & 0 & m_t
  \end{matrix}
  \right) \nonumber \\
  &=& \left(
\begin{array}{ccc}
\frac{g}{2M}\Upsilon_u \bar{\phi}_1 \alpha_1 |J'|^2 & 0 & 0 \\
0 &\frac{g}{2M} \Upsilon_u \bar{\phi}_1 \alpha_2 |I'|^2& 0 \\
0 & 0 &\frac{g}{2M} \Upsilon_u \bar{\phi}_1 \alpha_3 |K'|^2
\end{array}\right), \label{m.uct}
\end{eqnarray}

\begin{eqnarray}
& &  \left(
  \begin{matrix}
    m_d & 0 & 0 \\
    0 & m_s & 0 \\
    0 & 0 & m_b
  \end{matrix}
  \right) \nonumber \\
  &=& \left(
\begin{array}{ccc}
\frac{g}{2M}\Upsilon_d \phi_1 \alpha_1 |J'|^2 & 0 & 0 \\
0 & \frac{g}{2M}\Upsilon_d \phi_1 \alpha_2 |I'|^2 & 0 \\
0 & 0 & \frac{g}{2M}\Upsilon_d \phi_1 \alpha_3 |K'|^2
\end{array}
\right), \label{m.dsb}
\end{eqnarray}

\begin{eqnarray}
& &  \left(
  \begin{matrix}
    m_{\nu_e} & 0 & 0 \\
    0 & m_{\nu_\mu} & 0 \\
    0 & 0 & m_{\nu_\tau}
  \end{matrix}
  \right) \nonumber \\
  &=& \left(
\begin{array}{ccc}
\frac{g}{2M}\Upsilon_\nu \bar{\phi}_1 \beta_1 |J'|^2 & 0 & 0 \\
0 & \frac{g}{2M}\Upsilon_\nu \bar{\phi}_1 \beta_2 |I'|^2 & 0 \\
0 & 0 & \frac{g}{2M}\Upsilon_\nu \bar{\phi}_1 \beta_3 |K'|^2
\end{array}
\right), \label{m.nu1}
\end{eqnarray}

\begin{eqnarray}
& & \left(
  \begin{matrix}
    m_{e} & 0 & 0 \\
    0 & m_{\mu} & 0 \\
    0 & 0 & m_{\tau}
  \end{matrix}
  \right)  \nonumber \\
  &=& \left(
\begin{array}{ccc}
\frac{g}{2M}\Upsilon_e \phi_1 \beta_1 |J'|^2 & 0 & 0 \\
0 & \frac{g}{2M}\Upsilon_e \phi_1 \beta_2 |I'|^2 & 0 \\
0 & 0 & \frac{g}{2M}\Upsilon_e \phi_1 \beta_3 |K'|^2
\end{array}
\right). \label{m.e1}
\end{eqnarray}
The differences among masses of three fermion generations come from the factors $|\vartheta|^2$ and $\alpha_i$ (or $\beta_i$), with $i\in \{1,\,\, 2,\,\, 3\}$. Now we assume that the diagonal entries $\alpha_i$ (or $\beta_i$) are in the same order of magnitude. In other words, the relationship among three generation fermion masses are mainly determined by the relationship among $|\vartheta|$ in (\ref{IJK.<}) and (\ref{koide.ijk}). Ideally, one can let values of $\alpha_i$ (or $\beta_i$) be the same, and the mass relationships read
\begin{eqnarray}
  && m_1<m_2<m_3, \quad \text{with} \,\,\, \frac{m_2}{m_1}=e^{\theta_2}, \,\,\, \frac{m_3}{m_2}=e^{\theta_1} \label{m.<}, \\
  && m_1+m_2+m_3 =\frac{2}{3} (\sqrt{m_1}+\sqrt{m_2}+\sqrt{m_3})^2. \label{koide. m}
\end{eqnarray}
Here $m_i$ (with $i \in \{1,\,\, 2,\,\, 3\}$) denote the masses of three fermion generations in (\ref{m.uct}), (\ref{m.dsb}), (\ref{m.nu1}) and (\ref{m.e1}). The relationship (\ref{m.<}) means that the masses of three generations can get different values, and the differences among those masses are characterized by $\theta_1$ and $\theta_2$. The relationship (\ref{koide. m}) is the empirical Koide's relation~\cite{Koide:1982ax,Koide:1982si,Koide:1983qe,Li:2005rp,Li:2006et} of fermion masses. Now this empirical relation is a theoretical output of our model. We define
\begin{equation}  \label{koi}
  \Theta_{\mathrm{Koide}} = \frac{m_1+m_2+m_3}{\frac{2}{3}(\sqrt{m_1}+\sqrt{m_2}+\sqrt{m_3})^2} .
\end{equation}
Then, the values of $\Theta_{\mathrm{Koide}}$ calculated from physical data~\cite{pdg17} of leptons and quarks are show as follows:

\begin{center}
\begin{tabular}{cccc}
\toprule
& $e$, $\mu$, $\tau$& $u$, $c$, $t$ & $d$, $s$, $b$ \\
\midrule
$\Theta_{\mathrm{Koide}}$ & $0.999552 \pm 0.000024$ & $1.272 \pm 0.001$ & $1.094 + 0.005- 0.003$ \\
\bottomrule
\end{tabular}
\end{center}

\par
It is clear that the Koide's relation reflects the relationship among lepton masses with high precision, and that relation works, more or less, worse for quarks with heavier masses. In this work, since the relationship (\ref{koide.ijk}) is strictly satisfied in the ideal situation, how close is the $\Theta_{\mathrm{Koide}}$ to $1$ is determined by the magnitudes of differences among $\alpha_i$ for quarks (or $\beta_i$ for leptons). As shown in the above table, the differences among $\beta_i$ are small enough comparing with the Yukawa coupling coefficients of leptons and can be safely omitted. However, the differences among $\alpha_i$ can not be omitted without affecting the accuracy of Yukawa coupling coefficients. It is, in some sense, natural to speculate that the differences among $\alpha_i$ (or $\beta_i$) become more important with the corresponding masses getting heavier.

Thus, we construct the geometric paradigm of the standard model, with the $3$ fermion generations being the output of our combined extension. The mass relationships among those generations are determined by the adjustment on the imaginary units in our extension, and other parameters of mass matrices (such as the CKM and PMNS matrices) are hidden in $Q_2$ and $Q_1$.

Note that the order one condition is not used in recent works~\cite{Chamseddine:2013kza,Chamseddine:2013rta}. If the above extensions are made on the geometry without the order one condition, fermion generations can also emerge in a similar way as our work. In such geometry, the tensor product extension still contributes factors $Q$ on the action. With the quaternion extension, $Q$ are also rotation matrices on vectors $(L_J,L_J,L_K)$ of fermions, then the 3 fermion generations emerge. Moreover, when the action of geometry without order one condition truncates to low energy limit, the action reduces to that with the order one condition and is equivalent to our model.

We mention that there are several requirements taken along the way of our extensions. The noncommutative geometry is endowed with some essential features which are necessary for the geometry to reflect physical reality. One part of our requirements are the conditions which are fulfilled to keep the essential features of the ordinary noncommutative geometry unchanged in our combined extension:

(i) Scalar products on the fiber and the base spaces are respectively defined. This condition keeps the structures of the extension restricted to the fiber and the base spaces being linear structures. The condition is allowed by the different physical interpretations of mathematical structures of the fiber and the base spaces.

(ii)The extended algebra $\Delta A$ is restricted to $\Delta A^+$. This restriction ensures that the extended spectral triple is endowed with a $\mathbb{Z}/2-$grading. The restriction can come from the restriction of $\mathcal{H}_F$ to $\mathcal{H}_F^+$ in the original noncommutative geometry~\cite{nc.rev,Chamseddine:2014nxa}.

The other part of the requirements in our extensions is the inputs of the model to characterize the emergence of three generations and the mass relationships among those generations:

(i) The adjustments on the quaternion extension are made to reveal the connection between the quaternionic physical framework and the features (i.e., the generation number and the mass relationships among those generations) of fermions.

(ii) We let $|G|=1$ and $Q$ be diagonalizable matrices to simplify the quaternion extension on structures restricted to the base space.

Thus, the {\it ad hoc} fermion generation number and the mass parameters corresponding to the mass relationships among those fermion generations of the standard model are, in some sense, replaced by our extensions with the above conditions fulfilled. Since our work is just a toy model, we can not directly give the conclusion that our inputs are more ``economical'' than selecting {\it ad hoc} generation number and other mass parameters mentioned above. However, to understand the emergence of three fermion generations and corresponding mass relationships from one conceptual starting point (i.e., to make almost trivial extensions on one model to understand {\it ad hoc} inputs of that model) can still have enlightening value for the insight on the standard model and other theoretical frameworks, e.g., the noncommutative geometry, which couples the standard model with gravity.

\section{Conclusion}
\label{sec3}

In summary, we show the origin of $3$ fermion generations from noncommutative geometry. The framework of this geometry has physically trivial extensions, i.e., the tensor product and the quaternion extensions, in usual conditions. We consider the special case to combine these two extensions, and show that fermions naturally split into $3$ generations with right mass relationships among those generations. Our toy model still needs more conceptual understanding. For instance, we need the understanding of the nonlinear anticommutator. However, there is an essential advantage of our work. We give a way to find physical effects beyond a model by minimal, i.e., almost trivial, extensions on this model. As shown above, different mathematical structures can be physically equivalent in usual conditions, and the combination of several these structures can have nontrivial effects. Moreover, we show the possibility and impact of expressing physical framework by more complicated mathematical structures, e.g., tensor product structures and quaternion, to reflect the reality with exceptive features.

\section{Supplement: a simple example of the noncommutative geometry}

To make this work easier to read, we add an example of the application of the noncommutative geometry~\cite{Chamseddine:1991qh,nc.rev,Chamseddine:2006ep} to a U$(1)$ gauge symmetry as a supplement.

The fundamental structures of the ordinary noncommutative geometry~\cite{Chamseddine:1991qh,nc.rev,Chamseddine:2006ep} come from the product manifold $M\times F$ and the corresponding spectral triple $(\mathcal{A}, \mathcal{H}, D) = (\mathcal{A}_F, \mathcal{H}_F, D_F)\otimes (C^\infty (M), L^2(M,S), \partial\!\!\!/_M)$. To reflect the U$(1)$ gauge symmetry, the symplectic assumption and the nontrivial condition in the ordinary noncommutative geometry~\cite{concep,why.sm} should be removed. Then, in the minimal case, the algebra $\mathcal{A}_F$ in the spectral triple $(\mathcal{A}_F, \mathcal{H}_F, D_F)$ of the geometry is simply represented by ${\bf 1}$, i.e., the algebra over $\mathbb{C}$ as mentioned above. The Hilbert space $\mathcal{H}_F$ then reads $\mathcal{H}_F={\bf 1} \otimes {\bf 1}^0$. This Hilbert space has no sub-space endowed with a $\mathbb{Z}/2-$grading $\gamma_F$ because a $\mathbb{Z}/2-$grading asks for nontrivial structures of $\mathcal{H}_F$ such as ${\bf 2}\otimes {\bf 1}^0$. The representations of fermions are contained by the sub-space of $\mathcal{H}_F$ endowed with a $\mathbb{Z}/2-$grading. Thus, there is no fermion representation in the geometry of U$(1)$ gauge symmetry. We mention that the ordinary U$(1)$ gauge theory contains not only the U$(1)$ gauge bosons, but also all the U$(1)$ charged particles (fermions) under the U$(1)$ gauge transformation. This is because there are experimental data telling the ordinary U$(1)$ gauge theory what particles are involved in this theory. However, representations of fermions in the noncommutative geometry are theoretical outputs~\cite{concep,why.sm} of the geometry. In the minimal case without the symplectic assumption and the nontrivial condition, there is no output of fermions. Thus, to get the U$(1)$ noncommutative geometry including all U$(1)$ charged particles, one should construct the geometry with all the conditions in~\cite{concep,why.sm} fulfilled and get all the fermion outputs, and then select the U$(1)$ part from the geometry.

In the noncommutative geometry, bosons come from the inner fluctuation $A=\sum_i a_i[D, a_i']$ of the geometry, with $a_i,\,\, a_i' \in \mathcal{A}$. Here $D= \partial\!\!\!/_M \otimes 1 + \gamma_5\otimes D_F$, and $\mathcal{A}=\mathcal{A}_F\otimes C^\infty(M)$. As proved in the ordinary noncommutative geometry~\cite{Chamseddine:2010ud,Chamseddine:2013kza,Connes:1996gi,Connes:2006qj}, the discrete part $A^{(0,1)}=\sum_i a_i[D_F, a_i']$ of the inner fluctuation gives the representations of Higgs bosons, while the continuous part $A^{(1,0)}=\sum_i a_i[\partial\!\!\!/_M \otimes 1, a_i']$ of the inner fluctuation gives the representations of gauge bosons. Since $A_F={\bf 1}$, $A^{(0,1)}$ is governed by the U$(1)$ gauge transformation, and then $A^{(0,1)}$ gives the representations of U$(1)$ gauge bosons.

Applying the spectral action principle~\cite{Chamseddine:1996zu,Chamseddine:2006ep,Connes:1988ym} to the above spectral triple, one can directly get the action of the geometry. This action is formally the same as $S_b$ in (\ref{S.bi.1}), with the coupling terms of $G_{\mu\nu}$ and $B_{\mu\nu}$ removed. The gravity part of the action comes from the spectral triple $(C^\infty (M), L^2(M,S), \partial\!\!\!/_M)$ and the gauge coupling term comes from the spectral triple $({\bf 1}, {\bf 1}\otimes {\bf 1}^0, D_F)$. As discussed in this work, our combined extension on the noncommutative geometry only manifests itself in the fermion coupling terms of the action. Thus, the U$(1)$ noncommutative geometry is invariant under our extensions.

To represent fermions with right quantum numbers, there should be the symplectic assumption and the nontrivial condition~\cite{concep,why.sm} in the geometry, and then our combined extension can characterize the emergence of the three generations of fermions and the corresponding mass relationships among those generations..


\section*{Acknowledgments}

This work is supported by the National Natural Science Foundation of China (Grant No.~11475006).

\end{document}